\documentclass[superscriptaddress,twocolumn]{revtex4-1}

\usepackage[utf8]{inputenc}
\usepackage[english]{babel}

\usepackage{amsfonts,amsmath,amssymb}
\usepackage{float}
\usepackage{hyperref}
\usepackage{units}
\usepackage{graphicx}
\usepackage{upgreek}
\usepackage [autostyle, english = american]{csquotes}
\MakeOuterQuote{"}
\usepackage{placeins}
\usepackage{txfonts}

\begin{document}
	\title{Significant-loophole-free test of Bell's theorem with entangled photons}


\author{Marissa Giustina}\email[]{marissa.giustina@univie.ac.at}
	\affiliation{Institute for Quantum Optics and Quantum Information (IQOQI), Austrian Academy of Sciences, Boltzmanngasse 3, Vienna 1090, Austria.}
	\affiliation{Quantum Optics, Quantum Nanophysics, and Quantum Information, Faculty of Physics, University of Vienna,  Boltzmanngasse 5, Vienna 1090, Austria}

\author{Marijn A. M. Versteegh}
	\affiliation{Institute for Quantum Optics and Quantum Information (IQOQI), Austrian Academy of Sciences, Boltzmanngasse 3, Vienna 1090, Austria.}
	\affiliation{Quantum Optics, Quantum Nanophysics, and Quantum Information, Faculty of Physics, University of Vienna,  Boltzmanngasse 5, Vienna 1090, Austria}

\author{Sören Wengerowsky}
	\affiliation{Institute for Quantum Optics and Quantum Information (IQOQI), Austrian Academy of Sciences, Boltzmanngasse 3, Vienna 1090, Austria.}
	\affiliation{Quantum Optics, Quantum Nanophysics, and Quantum Information, Faculty of Physics, University of Vienna,  Boltzmanngasse 5, Vienna 1090, Austria}
	
\author{Johannes Handsteiner}
	\affiliation{Institute for Quantum Optics and Quantum Information (IQOQI), Austrian Academy of Sciences, Boltzmanngasse 3, Vienna 1090, Austria.}
	\affiliation{Quantum Optics, Quantum Nanophysics, and Quantum Information, Faculty of Physics, University of Vienna,  Boltzmanngasse 5, Vienna 1090, Austria}

\author{Armin Hochrainer }
	\affiliation{Institute for Quantum Optics and Quantum Information (IQOQI), Austrian Academy of Sciences, Boltzmanngasse 3, Vienna 1090, Austria.}
	\affiliation{Quantum Optics, Quantum Nanophysics, and Quantum Information, Faculty of Physics, University of Vienna,  Boltzmanngasse 5, Vienna 1090, Austria}
	
\author{Kevin Phelan}
	\affiliation{Institute for Quantum Optics and Quantum Information (IQOQI), Austrian Academy of Sciences, Boltzmanngasse 3, Vienna 1090, Austria.}

\author{Fabian Steinlechner}
	\affiliation{Institute for Quantum Optics and Quantum Information (IQOQI), Austrian Academy of Sciences, Boltzmanngasse 3, Vienna 1090, Austria.}


\author{Johannes Kofler}
	\affiliation{Max-Planck-Institute of Quantum Optics, Hans-Kopfermann-Straße 1, 85748 Garching, Germany}

\author{Jan-\AA ke Larsson}
	\affiliation{Institutionen för Systemteknik, Linköpings Universitet, 581 83 Linköping, Sweden}


\author{Carlos Abell\'{a}n}
	\affiliation{ICFO -- Institut de Ciencies Fotoniques, The Barcelona Institute of Science and Technology, 08860 Castelldefels, Barcelona, Spain}

\author{Waldimar Amaya}
	\affiliation{ICFO -- Institut de Ciencies Fotoniques, The Barcelona Institute of Science and Technology, 08860 Castelldefels, Barcelona, Spain}

\author{Valerio Pruneri}
	\affiliation{ICFO -- Institut de Ciencies Fotoniques, The Barcelona Institute of Science and Technology, 08860 Castelldefels, Barcelona, Spain}	
	\affiliation{ICREA -- Instituci\'{o} Catalana de Recerca i Estudis Avan\c{c}ats, 08015 Barcelona, Spain }

\author{Morgan W. Mitchell }
	\affiliation{ICFO -- Institut de Ciencies Fotoniques, The Barcelona Institute of Science and Technology, 08860 Castelldefels, Barcelona, Spain}	
	\affiliation{ICREA -- Instituci\'{o} Catalana de Recerca i Estudis Avan\c{c}ats, 08015 Barcelona, Spain }

		
\author{Jörn Beyer}
	\affiliation{Physikalisch-Technische Bundesanstalt, Abbestraße 1, 10587 Berlin, Germany }


\author{Thomas Gerrits}
	\affiliation{National Institute of Standards and Technology (NIST), 325 Broadway, Boulder, Colorado 80305, USA}

\author{Adriana E. Lita}
	\affiliation{National Institute of Standards and Technology (NIST), 325 Broadway, Boulder, Colorado 80305, USA}

\author{Lynden K. Shalm}
	\affiliation{National Institute of Standards and Technology (NIST), 325 Broadway, Boulder, Colorado 80305, USA}	

\author{Sae Woo Nam}
	\affiliation{National Institute of Standards and Technology (NIST), 325 Broadway, Boulder, Colorado 80305, USA}

\author{Thomas Scheidl}
	\affiliation{Institute for Quantum Optics and Quantum Information (IQOQI), Austrian Academy of Sciences, Boltzmanngasse 3, Vienna 1090, Austria.}
	\affiliation{Quantum Optics, Quantum Nanophysics, and Quantum Information, Faculty of Physics, University of Vienna,  Boltzmanngasse 5, Vienna 1090, Austria}

\author{Rupert Ursin}
	\affiliation{Institute for Quantum Optics and Quantum Information (IQOQI), Austrian Academy of Sciences, Boltzmanngasse 3, Vienna 1090, Austria.}

\author{Bernhard Wittmann}
	\affiliation{Institute for Quantum Optics and Quantum Information (IQOQI), Austrian Academy of Sciences, Boltzmanngasse 3, Vienna 1090, Austria.}
	\affiliation{Quantum Optics, Quantum Nanophysics, and Quantum Information, Faculty of Physics, University of Vienna,  Boltzmanngasse 5, Vienna 1090, Austria}
	
\author{Anton Zeilinger}\email[]{anton.zeilinger@univie.ac.at}
	\affiliation{Institute for Quantum Optics and Quantum Information (IQOQI), Austrian Academy of Sciences, Boltzmanngasse 3, Vienna 1090, Austria.}
	\affiliation{Quantum Optics, Quantum Nanophysics, and Quantum Information, Faculty of Physics, University of Vienna,  Boltzmanngasse 5, Vienna 1090, Austria}

	\date{\today}
	
	\begin{abstract}
	Local realism is the worldview in which physical properties of objects exist independently of measurement and where physical influences cannot travel faster than the speed of light. Bell's theorem states that this worldview is incompatible with the predictions of quantum mechanics, as is expressed in Bell's inequalities. Previous experiments convincingly supported the quantum predictions. Yet, every experiment requires assumptions that provide loopholes for a local realist explanation. 
	Here we report a Bell test that closes the most significant of these loopholes simultaneously. Using a well-optimized source of entangled photons, rapid setting generation, and highly efficient superconducting detectors, we observe a violation of a Bell inequality with high statistical significance. The purely statistical probability of our results to occur under local realism does not exceed $3.74 \times 10^{-31}$, corresponding to an 11.5 standard deviation effect.
\end{abstract}

	\maketitle
	
Einstein, Podolsky, and Rosen (EPR) argued that the quantum mechanical wave function is an incomplete description of physical reality \cite{EPR1935}. They started their discussion by noting that quantum mechanics predicts perfect correlations between the outcomes of measurements on two distant entangled particles. This is best discussed considering Bohm's example of two entangled spin-1/2 atoms \cite{Bohm1951,Bohm1957}, which are emitted from a single spin-0 molecule and distributed to two distant observers, now commonly referred to as Alice and Bob. By angular momentum conservation, the two spins are always found to be opposite. Alice measures the spin of atom 1 in a freely chosen direction. The result obtained  allows her to predict with certainty the outcome of Bob, should he measure atom 2 along the same direction. Since Alice could have chosen any possible direction and since there is no interaction between Alice and Bob anymore, one may conclude that the results of all possible measurements by Bob must have been predetermined. However, these predeterminate values did not enter the quantum mechanical description via the wave function. This is the essence of the argument by EPR that the quantum state is an incomplete description of physical reality \cite{EPR1935}. 

\begin{figure*}[!htp]
	\includegraphics[width=0.9\textwidth]{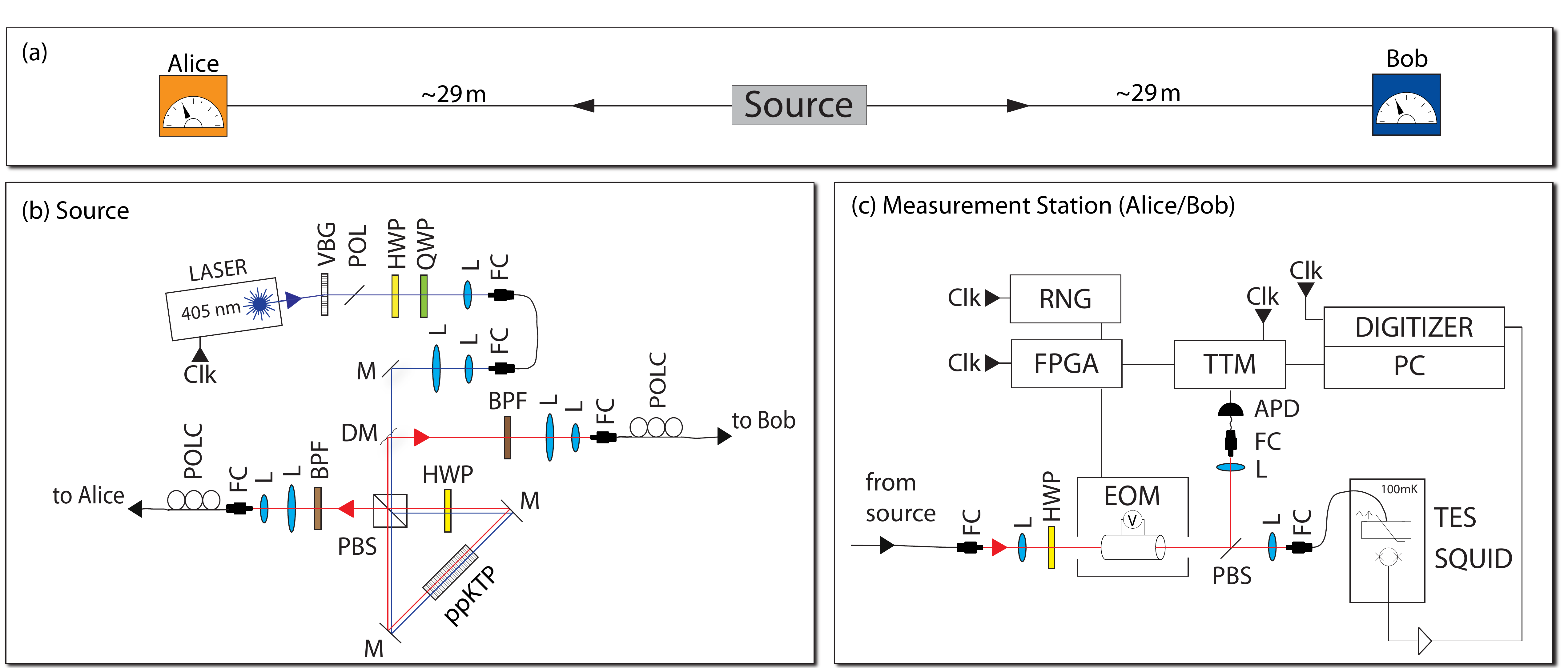}
	\caption{\label{fig:setup}(a) Schematic of the setup. (b) Source: The source distributed two polarization-entangled photons between the two identically constructed and spatially separated measurement stations \textit{Alice} and \textit{Bob} (distance $\unit[\approx58]{m}$), where the polarization was analyzed. It employed type-II spontaneous parametric down-conversion in a periodically poled crystal (\mbox{ppKTP}),  pumped with a $\unit[405]{nm}$ pulsed diode laser (pulse length:  $\unit[12]{ns}$ FWHM) at $\unit[1]{MHz}$ repetition rate. The laser light was filtered spectrally by a volume Bragg grating (VBG) (FWHM: $\unit[0.3]{nm}$) and spatially by a single-mode fiber.  The ppKTP crystal was pumped from both sides in a Sagnac configuration to create polarization entanglement. Each pair was split at the polarizing beam splitter (PBS) and collected into two different single-mode fibers leading to the measurement stations. 
	 (c) Measurement stations: 
	In each measurement station, one of two linear polarization directions was selected for measurement, as controlled by an electro-optical modulator (EOM), which acted as a switchable polarization rotator in front of a plate PBS. Customized electronics (FPGA) sampled the output of a random number generator (RNG) to trigger the switching of the EOM. The transmitted output of the plate PBS was coupled into a fiber and delivered to the TES. The signal of the TES was amplified by a SQUID and additional electronics, digitized, and recorded together with the setting choices on a local hard drive. The laser and all electronics related to switching/recording were synchronized with clock inputs (Clk). Abbreviations: APD: avalanche photodiode (see Fig.~\ref{fig:spacetime}); BPF: band-pass filter; DM: dichroic mirror; FC: fiber connector; HWP: half-wave plate; L: lens, POL: polarizer; M: mirror;  POLC: manual polarization controller;  QWP: quarter-wave plate; SQUID: superconducting quantum interference device; TES: transition-edge sensor;   TTM: time-tagging module.}
\end{figure*}

Bell's theorem states that quantum mechanics is incompatible with \textit{local realism}.  He showed that if we assume, in line with Einstein's theory of relativity, that there are no physical influences traveling faster than the speed of light (the assumption of \textit{locality}) and that objects have physical properties independent of measurement (the assumption of \textit{realism}), then correlations in measurement outcomes from two distant observers must necessarily obey an inequality  \cite{Bell1964}. Quantum mechanics, however, predicts  a violation of the inequality for the results of certain measurements on entangled particles. Thus, Bell’s inequality is a tool to rule out philosophical standpoints based on experimental results. Indeed, violations have been measured.

\begin{figure*}[htp]
	\includegraphics[width=0.9\linewidth]{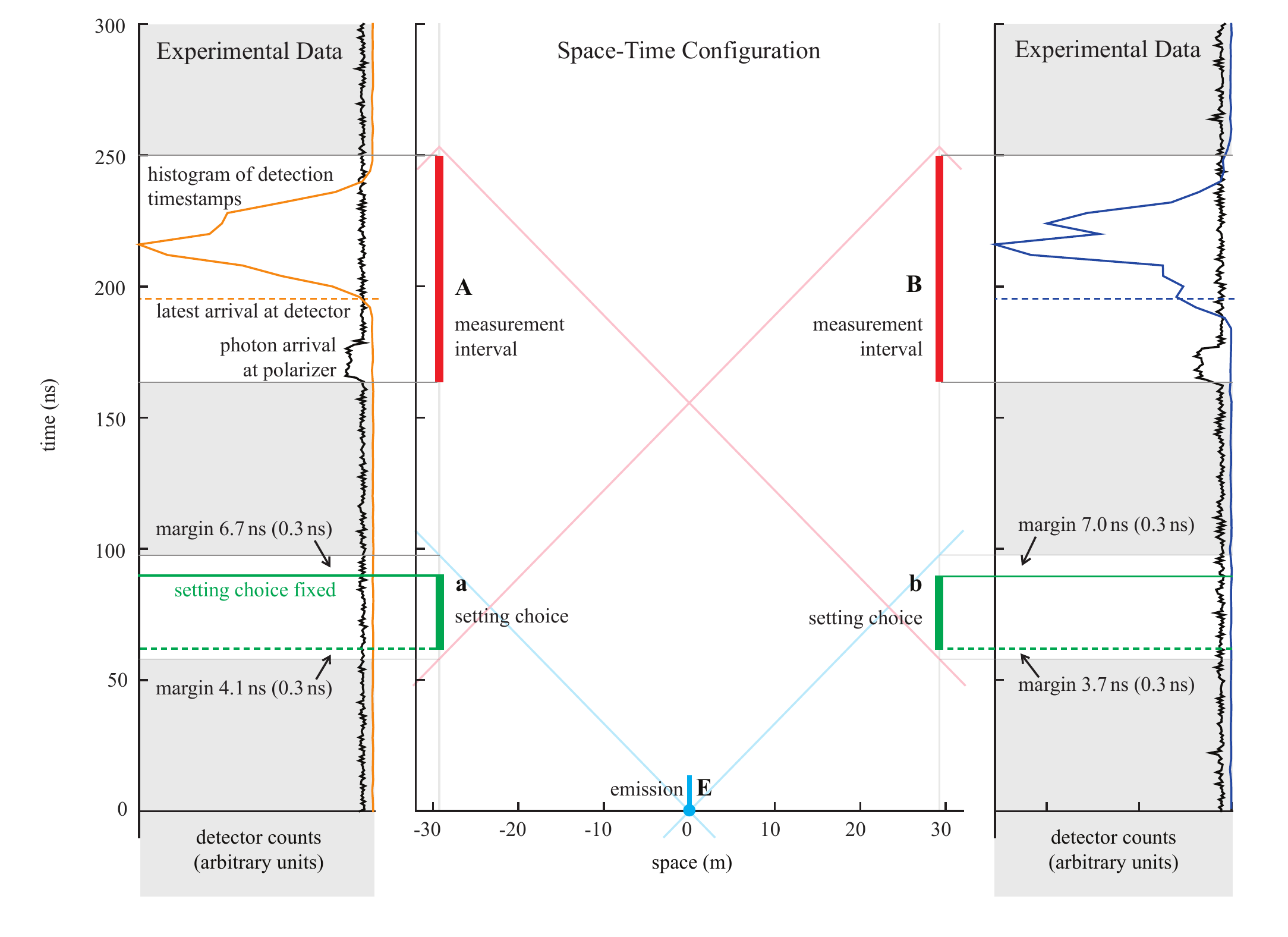}
	\caption{Space-time diagram representing experimental design and construction. The center block depicts to scale the approximate space-time configuration of our experiment. (Deviations from a purely one-dimensional construction are negligibly slight so that this diagram accurately characterizes our space-time layout.) The emission interval is represented in light blue ($\textbf{E}$), the selection of measurement  settings at Alice and Bob is confined to the green bars $\textbf{a}$ and $\textbf{b}$ respectively, and the measurement takes place within the red bars $\textbf{A}$ and $\textbf{B}$.
	The setting choice interval is constrained from the one side by the forward light cone of the earliest possible emission event and on the other side by the backward light cone of the end of the distant measurement interval. The diagonal lines indicate the speed of light in vacuum. The safety margins between each green bar and the relevant light cones were found by conservatively aggregating measurements of physical lengths and timing delays. The parenthesized values represent the combined standard deviation of the involved measurements, assuming independent and normal-distributed uncertainty. The narrow blocks to the left and the right of the center block depict experimental data at Alice and Bob respectively. Alice's and Bob's settings are selected by their random number generators (RNGs) at times indicated by the solid green horizontal lines. 
The orange and blue histograms each represent a distribution of photon detection times relative to the start of each trial. The orange and blue dotted lines represent the latest possible arrival time at the TES of photons created during the emission interval. The black histograms depict the arrival time of photons at the plate PBS, as characterized with a calibrated avalanche photodiode in the reflected output of the plate PBS (APD in Fig.~\ref{fig:setup}(c)).  \label{fig:spacetime}
	}
\end{figure*}

Do these experimental violations invalidate local realism? That is not the only logical possibility. The experimental tests of Bell’s inequality thus far required extra assumptions, and therefore left open \textit{loopholes}  that still allow, at least in principle, for a local realist explanation of the measured data. (Note that empirically closing a loophole might still require the validity of some specific assumptions about the experiment.)

The \textit{locality loophole} (or \textit{communication loophole}) is open if the setting choice or the measurement result of one side could be communicated to the other side in time to influence the measurement result there. Space-like separation of each local measurement from both the distant setting choice and the distant measurement closes the locality loophole. In an experiment, this can be ensured by independently choosing the measurement settings on both sides so quickly that no physical signal (limited by the speed of light) can pass information about the chosen setting or the measurement result to the other side in time to be relevant for the measurement there.

The \textit{freedom-of-choice loophole} refers to the requirement, formulated by Bell, that the setting choices are "free or random" \cite{Bell1990}. For instance, this would prohibit a possible interdependence between the choice of measurement settings and the  properties of the system being measured. Following  Bell, we describe all properties of the system with the variable $\lambda$, which represents "any number of hypothetical additional complementary variables needed to complete quantum mechanics in the way envisaged by EPR." \cite{Bell1990} 
This loophole can be closed only under specific assumptions about the origin of $\lambda$. Under the assumption that $\lambda$ is created with the particles to be measured, an experiment in which the settings are generated independently at the measurement stations and space-like separated from the creation of the particles closes the loophole.

The \textit{fair-sampling loophole} (or \textit{detection loophole}) refers to the following issue: It is conceivable 
under local realism that a sub-ensemble of emitted particles violates a Bell inequality, while the total ensemble does not. The loophole is exploited if an experiment detects only this sub-ensemble and assumes that it represents the entire ensemble \cite{Pearle1970}. It is possible to close the loophole by detecting the particles with adequate efficiency; the situation can be made even cleaner by using a Bell inequality that does not make such a "fair sampling" assumption in the first place.

There is a long history \cite{Brunner2014, LarssonLoopholes2014} of experimental Bell tests \cite{Freedman1972,Aspect1982,Weihs1998,Rowe2001,Matsukevich2008,Hofmann2012,Ansmann2009,Scheidl2010,Aguero2012,Giustina2013a,Christensen2013,Hensen2015} addressing individual loopholes,  though the long-term goal has always been to close all loopholes in a single experiment. To name a few, Aspect \textit{et al.}'s 1982 experiment \cite{Aspect1982} first employed rapid switching of the measurement settings; Weihs \textit{et al.}'s 1998 experiment \cite{Weihs1998} improved this with fast random switching; Scheidl \textit{et al.}\ \cite{Scheidl2010} addressed  freedom-of-choice in 2010 while also closing the locality loophole; Rowe \textit{et al.}\ \cite{Rowe2001} were first to close the fair-sampling loophole in 2001 and were followed by several experiments in a variety of systems \cite{Matsukevich2008,Ansmann2009,Giustina2013a,Christensen2013,Hensen2015}. 
It has only recently become possible to address all three aforementioned loopholes in a single experiment \cite{Hensen2015,NIST2015} (where \cite{Hensen2015} reported a \textit{p}-value of 0.039). 
Here we report the violation of a Bell inequality using polarization-entangled photon pairs, high-efficiency detectors, and fast random basis choices space-like separated from both the photon generation and the remote detection. We simultaneously close all three aforementioned loopholes in a single experiment with high statistical significance and thus provide strong support for the idea that nature cannot be described within the framework of local realism.

The experimental setup, located in the sub-basement of the Vienna Hofburg castle, is illustrated in Fig.~\ref{fig:setup}(a). Our pair source [Fig.~\ref{fig:setup}(b)] used spontaneous parametric down-conversion (SPDC) in a periodically poled nonlinear crystal and generated high-quality entanglement using a Sagnac configuration \cite{fedrizzi2007}. We optimized the focusing parameters of the pump and down-conversion collection modes for high heralding efficiency \cite{Bennink2010,Steinlechner2015}. The photons were coupled into single-mode fibers and distributed to two distant measurement stations "Alice" and "Bob" [Fig.~\ref{fig:setup}(c)] where polarization measurements were performed in one of two setting angles per side. While the photons were in flight, the choice of measurement setting was made in each station by a random number generator (RNG) \cite{Abellan2014,Abellan2015} situated there. The measurement was implemented by a fast electro-optical modulator (EOM) followed by a polarizer and a transition-edge sensor (TES) single-photon detector \cite{Lita2008}. The signal from the TES was amplified by a series of superconducting \cite{Drung2007} and room-temperature amplifiers, digitized, and recorded locally on a hard drive. In addition, each implemented setting was recorded locally at each measurement station using a time-tagging module. The photon and setting data stored locally in the measurement stations were collected by a separate computer that evaluated the Bell inequality.

To close the freedom-of-choice and locality loopholes, a specific space-time configuration of the experiment was chosen, as depicted in the center plot of Fig.~\ref{fig:spacetime}. 
As discussed before, it was necessary to space-like separate each local setting choice  (green bars labeled $\textbf{a}$ and $\textbf{b}$) from the measurement on the other side (red bars $\textbf{A}$ and $\textbf{B}$), as well as from the photon emission (blue bar $\textbf{E}$),  which we consider to be the origin of $\lambda$.

To ensure synchronized timing throughout the experiment, we locked the RNG, TTM (time-tagging module), and digitizer to a $\unit[10]{MHz}$ master oscillator. A $\unit[1]{MHz}$ clock, phase-synchronized to this master oscillator, regulated the laser pulsing and switching of EOMs. To confirm the space-time configuration in our experiment, we precisely characterized the delays of all relevant electrical and optical signals relative to this clock using an oscilloscope and a fast photodiode. In particular, we characterized three events in the experiment:
\begin{enumerate}

 \item \emph{Emission:} The origin of the space-time diagram, indicated as a blue dot, represents the earliest possible photon emission. This point corresponds to the leading edge of the pump laser pulse reaching the SPDC crystal in the source. The length of the blue bar $\textbf{E}$ indicates the pulse duration of the pump laser. 

\item \emph{Setting choice:} We allow approximately $\unit[26]{ns}$ for the RNG to produce and deliver a setting choice by generating four raw bits and computing their parity. The randomness in each raw bit is derived from the phase, randomized by spontaneous emission, of an optical pulse. These $\unit[26]{ns}$ include a creation and throughput time of approximately $\unit[11]{ns}$ for one raw bit and an additional $\unit[15]{ns}$ for three additional bits. As described in the supplemental material \cite{supplemental_material} \nocite{RukhinNIST2010,LEcuyerACM2007,Cabrera1998,Lita2010,GiustinaDiss,Bednorz2015}, this reduces the chance of predicting the settings to $\varepsilon_{\rm RNG} \leq 2.4 \times 10^{-4}$ \cite{Abellan2014,Abellan2015}. The solid green horizontal lines in the space-time diagram indicate the latest possible time at which the random phase was sampled inside the respective RNGs for use in a setting choice, while the dashed green lines indicate the earliest possible random phase creation for the first (of the four) contributing raw bit. The configuration ensures conservatively estimated margins of $\approx\unit[4]{ns}$ for the space-like separation of each setting from the distant measurement and $\approx\unit[7]{ns}$ for the space-like separation of each setting from the emission event (see Fig.~\ref{fig:spacetime} for more detail, including error estimates).

	\item \emph{Measurement:} After a photon pair is emitted by the crystal, the photons are coupled into two single-mode optical fibers that direct one photon each to Alice’s and Bob’s distant locations. At each measurement station the photons are coupled out of fiber and sent through an EOM and a polarizer that transmit a particular polarization based on the setting choice from the RNG. The photons transmitted through the polarizer are coupled back into optical fiber (SMF-28) and sent to the TES. For monitoring purposes, we use an avalanche photodiode to detect the photons that are reflected from the polarizer (black histograms in Fig.~\ref{fig:spacetime}). Using the arrival time information from this monitoring port, and assuming photons travel at the speed of light in their respective media, we infer that the latest time a photon could arrive at a TES after being emitted from the source is approximately $\unit[195]{ns}$. This is represented by the dashed orange and blue lines on the space-time diagram in Fig.~\ref{fig:spacetime}. 

After a photon is absorbed by the TES, the resulting detection signal is read out using a  SQUID sensor, which introduces jitter into the signal. This electrical signal then travels through cables until it reaches a digitizer (the signals take approximately $\unit[64.4]{ns}$ and $\unit[65.5]{ns}$ to travel from the TES to Alice’s and Bob’s digitizers, respectively). Because the shape of the readout signal depends on the energy of the photons absorbed by the TES, the shape can be used to distinguish both unwanted background light (primarily blackbody photons) and excess noise from the $\unit[810]{nm}$ photons produced by the source.  We therefore use the digitizer to record the profiles of these amplified TES pulses.  When the amplified signal from the TES crosses a voltage threshold (around 55\% of the expected height of an $\unit[810]{nm}$ photon determined from calibration data), the signal is saved by the data acquisition system for further processing. During the analysis, if the recorded trace crosses a voltage level fixed at around 75\% of the expected pulse height from an $\unit[810]{nm}$ photon, then it is considered to be a detection event. This level was chosen to eliminate with near certainty lower-energy blackbody photons. The time that the trace crosses a level set at around 20\% of the expected pulse height is used to timestamp the detection event. We consider the detection event to be complete and the outcome fixed by this point. Histograms of these detection times relative to the start of the trial are shown in orange and blue in Fig.~\ref{fig:spacetime}. After accounting for cable delays, all events that fall inside the measurement windows \textbf{A} and \textbf{B} are ensured to be space-like separated from the relevant setting choice at the other party.

\end{enumerate}

Closure of the fair-sampling loophole does not rely on space-time considerations and can be observed in the experimental data. The Clauser-Horne (CH) \cite{Clauser1974} or the Eberhard \cite{Eberhard1993} inequality can be derived without the fair-sampling assumption. These inequalities can be violated with system heralding efficiencies larger than 2/3. We employed a CH-Eberhard (CH-E) type inequality, which requires only one detector per side and restricts the probabilities of outcomes---"+" for a detection and "0" for no detection---in the following way \cite{Kofler2015,Bierhorst2015}:
\begin{equation}\label{eq:eberhardineq}
J\equiv p_{++}(a_1b_1)-p_{+0}(a_1b_2)-p_{0+}(a_2b_1)-p_{++}(a_2b_2) \leq 0.
\end{equation} 
In every trial, Alice chooses setting $a_1$ or $a_2$, and Bob chooses $b_1$ or $b_2$. They write down their respective outcomes "+" or "0". Combining their data at the end of the experiment, they estimate the  probabilities that appear in the inequality. E.g., $p_{+0}(a_1b_2)$ is the probability that, conditioned on the setting choices $a_1$ and $b_2$  for a given trial, Alice observes a detection event and Bob registers no detection. Our experiment employed locally defined time slots and was thus also not vulnerable to the coincidence-time loophole \cite{larsson2004,Larsson2014}.

The inequality can be violated using Eberhard states of the form \cite{Eberhard1993}
\begin{equation}
	|\Psi\rangle = \frac{1}{\sqrt{1+r^2}}\left(|{\rm V}\rangle_{\rm A } |{ \rm H}\rangle_{\rm B} + r|{\rm H}\rangle_{\rm A} |{\rm V}\rangle_{\rm B} \right) \label{eq:state}
\end{equation}
where H and V  are horizontal and vertical polarizations and the subscripts A and B indicate Alice's and Bob's photons respectively.

\begin{figure}[t]
	\includegraphics{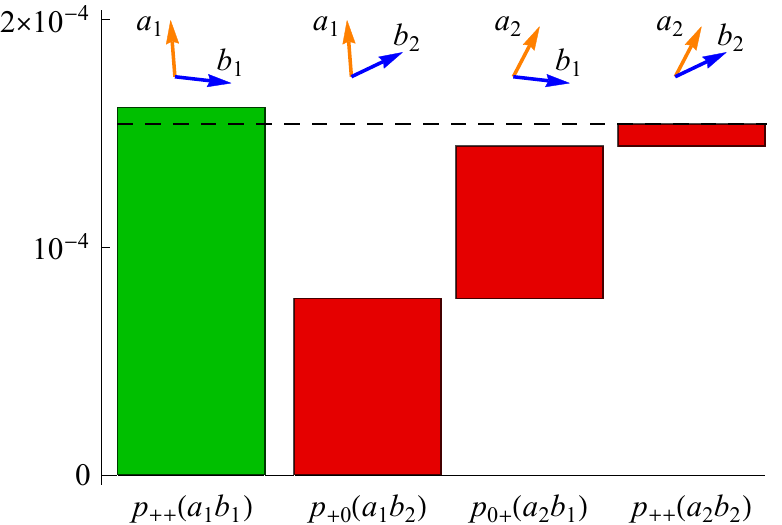}
	\caption{Bar chart of the four joint probabilities entering the Bell  inequality (Eq. \ref{eq:eberhardineq}). Since the green bar representing $p_{++}(a_1 b_1)$  outweighs the sum of the other three red bars, the $J$-value is positive and the CH-Eberhard inequality is violated.  \label{fig:histogram}}
\end{figure}

The optimal values for $r$ and the setting angles depend on the performance of the setup and can be estimated using a quantum mechanical model \cite{Kofler2013}. We characterized the system using the product state ($r = 0$) and the maximally entangled state ($r = -1$). We found visibilities of over 99\% in both the H/V and diagonal bases and system heralding efficiencies of approximately 78.6\% in the Alice arm and approximately 76.2\% in the Bob arm. These efficiencies represent a ratio of two-fold coincidence events divided by singles counts (i.e.\ total events measured in one detector) directly measured over the entire system and not corrected for any losses. We set a state with $r\approx-2.9$ and measured at angles $a_1=94.4^{\circ}$, $a_2=62.4^{\circ}$, $b_1=-6.5^{\circ}$, and $b_2=25.5^{\circ}$ for approximately 3\,510 seconds, and obtained the  probabilities shown in Fig.~\ref{fig:histogram}, corresponding to a $J$-value of $7.27 \times 10^{-6}$. For the pure state (\ref{eq:state}), the above mentioned detection efficiencies, and 3\,500 down-conversion pairs produced per second (see supplemental material), quantum mechanics predicts an optimal $J$-value of about $4 \times 10^{-5}$ \cite{Kofler2013}. That the measured value is smaller can be explained mostly by non-unity state visibility and nonzero background.

We compute the statistical significance of our measured violation under full experimental memory \cite{Barrett2002,Gill2003_1,Gill2003_2}, without assuming independent and identically distributed (IID) experimental trials \cite{Kofler2015}. We also account for the excess  predictability of the random setting choices and find that under local realism, the probability of observing our measured $J$-value does not exceed a \textit{p}-value of $3.74\times 10^{-31}$ (see supplemental material). Our analysis uses neither Gaussian approximation nor the IID assumption, but for comparison, for a large-sample experiment that allows these two, an 11.5 sigma violation gives this \textit{p}-value. In light of such an exceedingly small \textit{p}-value, we remark that the confidence in the experiment as a whole is limited not by the statistical strength of the violation but rather by other more general errors, which might happen in any experiment and could for example be systematic, human, or come from other limitations of the apparatus.

Our experiment showed a strong violation of local realism using exacting experimental technique and rigorous statistical analysis. Employing state of the art random number generators, we space-like separated the setting choices, measurements, and emission event to close the locality and freedom-of-choice loopholes simultaneously. We achieved high system heralding efficiencies and closed the fair-sampling loophole as well. In addition, we closed the coincidence-time loophole in our experiment by using locally-defined time slots. We closed the memory loophole by computing the statistical significance of the violation without assuming independently and identically distributed experimental trials. Our experiment provides the strongest support to date for the viewpoint that local realism is untenable.

By closing the freedom-of-choice loophole to one natural stopping point---the first moment at which the particles come into existence---we reduce the possible local-realist explanations to truly exotic hypotheses.
Any theory seeking to explain our result by exploiting this loophole would require $\lambda$ to originate before the emission event and to influence setting choices derived from spontaneous emission.
It has been suggested that setting choices determined by events from distant cosmological sources could push this limit back by billions of years \cite{Gallicchio2014}. 

\vspace{2ex}
We thank Reinhold Sahl and the Vienna Hofburg for the use of their sub-basement; in particular we thank Michael Bamberger, Angelika Stephanides, Wolfgang Weiland, and the night shift security officers. We acknowledge Scott Glancy, Sven Ramelow, Johannes Steurer, Max Tillmann, and Witlef Wieczorek for technical assistance and helpful discussions. MG acknowledges support by the program CoQuS of the FWF (Austrian Science Fund).
JK thanks Lucas Clemente for help with implementing data analysis code and acknowledges support from the EU Integrated Project SIQS. CA, WA, VP, MWM acknowledge the European Research Council project AQUMET, European Union Project QUIC (Grant Agreement 641122), Spanish MINECO under the Severo Ochoa programme (Grant. No.  SEV-2015-0522) and projects MAGO (Grant No. FIS2011- 23520) and EPEC (Grant No. FIS2014-62181-EXP), Catalan AGAUR 2014 SGR Grants No. 1295 and 1623, the European Regional Development Fund (FEDER) Grant No. TEC2013-46168-R, and by Fundaci\'o Privada CELLEX. This work was also supported by the NIST Quantum Information Science Initiative. This project was supported by the Austrian Academy of Sciences ({\"O}AW), the European Research Council (SIQS Grant No. 600645 EU-FP7-ICT), and the Austrian Science Fund (FWF) with SFB F40 (FOQUS).

\bibliographystyle{apsrev4-1.bst}
\bibliography{loophole-free_belltest1.bib}

\end{document}